\documentclass[twocolumn,twoside,slac_two]{revtex4}
\usepackage{graphicx}
\usepackage{fancyhdr}
\def\ls{{_<\atop^{\sim}}}
\def\gs{{_>\atop^{\sim}}}

\begin{document}

\title{Gamma-ray Burst Theory in the Fermi Era}

%

\author{Dafne Guetta}
\affiliation{Department of Physics and Optical Engineering, ORT Braude,
P.O. Box 78, Carmiel, Israel\\
INAF-Osservatorio astronomico di Roma, v. Frascati 33, 00040
Monte Porzio Catone, Italy}

\begin{abstract}
The Large Area Telescope (LAT) on-board the Fermi satellite detected emission above 20 MeV only in a small fraction of the long gamma-ray bursts (GRBs) detected by the Fermi Gamma-ray Burst Monitor (GBM) at 8 keV-40 MeV. Those bursts that were detected by the LAT were among the brightest GBM bursts. The emission detected by LAT  seerms to be delayed respect to the one detected by the GBM. In this review I will show the main implications of these LAT observations on the GRB models and discuss the importance 
of a synergy between Fermi and other telescopes.
\end{abstract}

\maketitle

\thispagestyle{fancy}


\section{Introduction}
Gamma ray bursts (GRBs) are one of the most powerful events in the universe. The
high energy photons emitted travel from cosmological distance tracing the star formation
history in the universe. More than 5 GRBs have been found at $z>4$  indicating that GRBs are the
most distant objects in the Universe.
Several papers have shown how GRBs evolve with redshift and have investigated the possibility to use these sources to probe
the Star Formation Rate at high redshift\cite{GP07}.

 While it is widely accepted that
GRBs are produced by the dissipation of energy in highly relativistic
winds driven by compact objects (see, e.g.,\cite{Msz06,P04,W03}
 for reviews) the physics of
wind generation and radiation production is not yet understood.
It is not known, for example, whether the wind luminosity is carried,
as commonly assumed, by kinetic energy or by Poynting
flux (e.g.\cite{DS02,L03} ), whether the radiating particles are accelerated by the dissipation
of magnetic flux or by internal shocks dissipating kinetic
energy, and whether the emission is dominated by synchrotron
or inverse-Compton radiation of accelerated electrons, as commonly
assumed, or by hadronic energy loss of accelerated protons
(see\cite{DF08}, and references therein).

Before Fermi, GRBs were detected by instruments sensitive mainly in the
100 to 1000 keV range like BATSE\cite{P99} and
the GRBM on BeppoSAX\cite{G04}. Measurements
at high energy were limited by the lower sensitivity and/or
the smaller field of view of higher energy instruments (e.g.
CGRO/EGRET-\cite{D95,H94}; AGILE/GRID-\cite{M09,G10}). 
The improved high energy, $\gs 1$ GeV, sensitivity and field
of view of the instruments on board the Fermi satellite\cite{At09,Band09} were expected to improve the quantity
and quality of high energy GRB data. Several GRBs were expected to be detected
in the $\sim$ GeV range by LAT.

However, one of the key results of Fermi is that the vast majority of GRBs detected by GBM were not detected
by LAT. This result implies qualitatively new constraints on GRB models.  
In an earlier study, Guetta et al. (2011)\cite{G11} have used the non-detection of GeV emission from
the majority of Fermi GRBs, to put upper limits on the GeV fluence of long GRBs. They find an
upper limit on the LAT/GBM fluence ratio of less than unity for 60\% of GRBs. In a later 
work of Beniamini et al (2011)\cite{Ben11} found an average upper limit to the fluence ratio of 0.13 during the prompt phase 
for the most luminous GRBs detected by the GBM but not by LAT. 
It is important to notice that putting upper limits on the LAT/GBM fluence ratio is crucial in order
to test the nature of the spectrum at high energies. This, in turn, will further constrain the emitting mechanism. I will discuss the results of these works on the LAT upper limits and implications in Section 3 of this review.

Another key result of Fermi is that the $\sim$ GeV emission detected by LAT seems to be  systematically delayed with respect
to the emission observed with the GBM at hundreds-of-keV energies (Fig. 2, left of Piron et al 2012\cite{Piron11})
Several physical models have been introduced to explain this delay. The most accredited are
the Afterglow Model: In this model the delay is interpreted as due to different origins of the low and high energy emissions,
the prompt low energy emission coming from the internal shocks while the high energy delayed emission
may be associated to the afterglow emission\cite{Ghir10}.
The other model is the Lorentz Invariance Violation: If the high energy photons are emitted during the prompt phase then the delay may be due to Lorentz Invariance Violation\cite{Amelino98}.

In this review I will discuss the possibility that the delay is due to Lorentz Invariance Violation and
show the implications of this scenario.
A key test of Lorentz invariance is an energy-dependent dispersion
effect, the possible variation of photon speed with energy.
This leads to a variation in photon arrival time with energy,  roughly given by 
\begin{equation}
\Delta t  \simeq  \frac{\Delta E}{M_{QG}} L ~,
\label{delaySMALLz1}
\end{equation}
which could be as large as seconds to hours for photons in the
$GeV$ to $TeV$ range if the distance $L$ travelled is
cosmological. Where  $M_{QG}$ is the quantum gravity mass thought to be
on the order of  $M_{Planck}$.
As we can see from this equation the effect is expected to be extremely small, however
for propagation over cosmological
distances (as GRBs are), the strong suppression by the Planck scale can
be compensated by the very large propagation times,
greatly amplyfing the obsevability of these tiny effects.
Several works have already reported limits on the
characteristic scale $M_{QG}$  with the desirable “Planck-
scale significance”, using Fermi data for several GRBs
like GRBs080916c, GRB090510, GRB090902B and
GRB090926A \cite{Fermi1,Fermi2,Fermi3}. In section 4 of this review I will 
summirize these results and will  propose a different strategy to put 
constraint on the  $M_{QG}$  which uses
the full GRB energy distribution, i.e. the spectrum, and
its temporal variations\cite{Fiore13}.

In section 5 we will discuss the necessity of a sinergy between the Fermi telescope, neutrino telescopes (like present IceCube and future KM3NET) and gravitational
wave detectors (like Virgo and LIGO). The sinergy with these detectors is fundamental as it will allow to put constraints on the hadronic emission models (Icecube) and to the progenitor models for the short Gamma-Ray Bursts.
In section 6 we report our conclusions.
\section{Unquestionable facts and Open Questions on  GRBs}
After  twenty years of extensive research in the Gamma Ray Burst field there
are few basic, unquestionable facts, which are common to
all GRBs and should be addressed by any theoretical
model. It is firm today that GRBs are:
\begin{itemize}
\item At cosmological distance, as they are typically observed at redshift $z>1$
\item The jet expands relativistically: high Lorentz factor $\Gamma\sim 100$ is required by observation
of high energy photons. This has solid confirmation
by the existence of afterglow emission,
which follows the interaction of the relativistic
ejecta with the ambient medium.
\item The energy released is up to few times the rest mass of Sun (if isotropic) in a few seconds.
\item There are two populations of GRBs separated
by their duration and hardness: short ($T90<2$ s) and long  ($T90>2$ s) GRBs 
\item The spectrum is not thermal and in the
vast majority of bursts, it has a broken power
law shape (the “Band” function, named after
the late David Band), peaking at sub-MeV, with
a fairly sharp decline at higher energies). 
\end{itemize}
Together with these unquestionable facts there
are still several fundamental open questions that need to be answered.
One of them is the composition of the jet: are hadrons present in the jet and what is their role in the emission mechanism?
Related to that we do not know yet  what are the main radiative processes, and physical explanation to
the broad band spectrum observed.
 The nature of the dissipation mechanism that leads to the emission of $\gamma$-rays is still poorly understood.
Regarding the progenitor models there are evidence that long GRBs are associated to the 
collapse of massive star and are connected to Supernova\cite{Hjort2003,GDV2007} while for short GRBs indirect evidence suggests
that short GRBs originate from binary mergers
\cite{Berger2005}, but there is no conclusive evidence yet\cite{Nys2009,Virg2011}. The finall product of both long and short GRB progenitors is probably a  Black Hole.
In this review I will show how the results of Fermi and the synergy between Fermi and neutrino and gravitational
wave telescopes can help us in addressing these open questions allowing a deeper understanding of the nature of GRBs.
In particular I will concentrate on two LAT key results: The paucity of GRBs detected by LAT and the delay between the 
GBM and LAT emission.
\section{Key result 1: Fermi upper limits}
In an earlier study, Guetta et al. (2011\cite{G11}) have used the non-detection of GeV emission from
the majority of Fermi GRBs, to put upper limits on the GeV fluence of long GRBs. They find an
upper limit on the LAT/GBM fluence ratio of less than unity for 60\% of GRBs. This is consistent
but not better than the EGRET-derived limits\cite{Ando2008}. In Beniamini et al 2011\cite{Ben11}, the authors use a more
subtle approach to further constrain these limits. As the LAT detected GRBs are also among the
brightest GRBs in the GBM band, they examine the brightest group of GBM
bursts with no LAT detection. They have analyzed the group of 18 most luminous GBM bursts, as these are expected to have the highest (undetected) GeV component. For these bursts, they obtain
upper limits on the GeV fluence. 
Fig. 1 depicts comparisons of the LAT/GBM fluence ratios of LAT bursts with the upper
limits on the LAT/GBM fluence ratio of LAT non-detected bursts for T = T90 sec.
The average LAT/GBM fluence ratio for LAT detected bursts is 0.09  for T90.
These limits are somewhat lower than the corresponding upper limit on this ratio derived for
 all 18 GBM bursts that is on average 0.13 during the prompt phase (T90). 
Fig. 1 shows that these limits are almost
uniform for all GBM bursts and do not seem to depend on the GBM fluence of the bursts. This
means that the upper limits derived here are mainly representative of the LAT detection limit and do not show any evidence for actual GeV signals in the GBM sample.
This LAT/GBM fluence ratio strongly constrains various emission models and rule out  any model in which there is a strong GeV component in the prompt mission.
In particular they limit strongly SSC models for the prompt emission as those will suggest a
second SSC component at the GeV\cite{Ando2008,Piran2009}. They pose a strong limit
on the conditions within the emitting regions showing no IC GeV Peak.
For the majority of the GBM and the LAT bursts the GeV fluence is compatible with the Band
extrapolation of the MeV emission. However, out of the LAT bursts, in three cases the Band extrapolation
of the MeV emission is higher than the observed fluence. Similarly the Band extrapolation
is higher than the LAT upper limits in 5 out of the 18 GBM bursts. Both results are consistent and
suggest that in some bursts we observe a decline in the spectral high energy slope between the
MeV and the GeV. This may be the first indication for the long sought after pair opacity break
in the high energy spectrum. If so it can enable us to estimate corresponding values of the bulk
Lorentz factors which are around a few hundred.
Another explanation may be that the electron energy distribution does not follow a power law over a
wide energy range\cite{G11}.
\begin{figure}
\includegraphics[width=75mm,height=65mm]{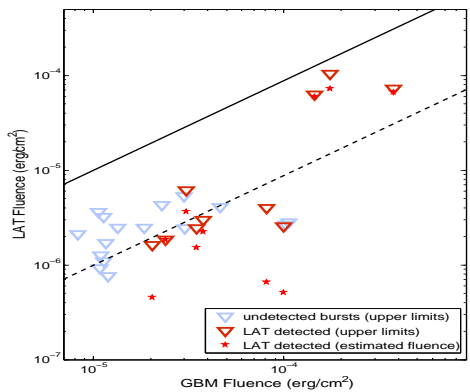}
\caption{The LAT fluence vs. the GBM fluence for two types of bursts: LAT undetected (the GBM sample) 
and LAT detected (the LAT sample)
bursts, at T90 duration. For the GBM bursts upper limits at a 90
Beniamini et al 2012 provide both  upper
limits and estimates of the fluence. The solid line marks equal fluence in the LAT and GBM bands, and the dotted line marks a LAT/GBM fluence ratio of 0.1.\cite{Ben11}}
\label{l2ea4-f1}
\end{figure}
\section{Key result 2: Constraints on the Lorentz Invariance Violation effect}
The candidate quantum-gravity effects that can be most
effectively studied using GRBs are those amounting to a violation\cite{Amelino98,Alfaro2000} or
deformation of Lorentz Invariance, induced by the Planck scale\cite{Amelino2002,Smolin2002}
($E_{\rm Plank}=M_{\rm Plank} c^2= 1.22\times 10^{19}$ GeV). In
particular it has been observed\cite{Amelino98}
that quantum properties of spacetime could produce an effective
energy-dependent dispersion effect, effectively introducing a small
dependence on energy of the speed of photons.  For propagation over
cosmological distances, the strong suppression by the Planck scale can
be compensated by the very large propagation times, greatly amplyfing
the obsevability of these tiny effects. Because of the smallness of
the effects at stake, it is important to find the best indicators of
the conjectured dispersion effect, also taking into account that we
have no control (and only a limited understanding) of the emitter of
the GRB signals we are interested in timing. Some ways to do this,
based on characteristic features in the GRBs lightcurves at high
energy, have already been proposed\cite{Amelino98,Fermi1,Fermi2,Fermi3,Ghir10}.
We here observe that energy-dependent dispersion can be also revealed
looking at the GRB spectral variability, and we argue that this
indicator could play a crucial role in building a case for (or
against) in-vacuo quantum-gravity-induced dispersion, as the data set
on GRBs observed at high energies keeps growing.
Several works have already reported limits on the characteristic scale
$M_{QG}$ with the desirable ``Planck-scale significance", using Fermi
data for several GRBs like GRBs080916c, GRB090510, GRB090902B and
GRB090926A\cite{Fermi1,Fermi2,Fermi3,Fermi4}. 
Most of these limits are obtained considering a single high energy
photon exept for the case GRB090510 where the Fermi collaboration\cite{Fermi3} establishes an interesting  bound  
$M_{QG}$  using the DisCan
method, { \it i.e.}  searching for time delays within the LAT data,
using the energy range 35 MeV-31 GeV and in the burst interval where
there is the most intense emission (0.5-1.45 s).  They find $|\Delta
t/\Delta E|<30$ ms GeV$^{-1}$ implying $M_{QG}/M_{Plank}>1.2$. 

The approach to use a single high energy photon has two main drawbacks: First, it is not
possible to know the exact time of emission of a single photon and
therefore these limits are intrinsically fuzzy. Second, for the same
reason it is difficult to assess the statistical meaning of each
limit, i.e. to which confidence interval it is referred to. To
overcome these problems (Fiore et al. 2013\cite{Fiore13})  propose a different strategy which uses the
full GRB energy distribution, i.e. the spectrum, and its temporal
variations. 
{\it The Method:} Energy-dependent dispersion effects can induce significant spectral
variations during the development of the GRB and its afterglow. Thus,
the accurate measure of spectral variations can put constraints to these
energy-dependent dispersion effects. There are two main problems with this
approach. The first is that dispersion induced spectral variations will
combine with possible intrinsic GRB spectral variations. Disentangling
the two in one single event is clearly extremely difficult. However,
intrinsic spectral variation will be independent on the GRB distance,
while dispersion induced variations will increase with the distance.
Therefore, the measure of significant spectral variability in a large
sample of bursts at different distances can put constraints on
energy-dependent dispersion effects or even provide a measure of the QG
energy scale. The second problem is that even in absence of intrinsic
spectral variations, dispersion induced spectral variations may be extremely
complex, because they depend on the temporal structure of the GRB. A
viable solution to this problem is to consider only clean events,
i.e. the first well defined peaks in the GRB evolution.
Fiore et al. 2013\cite{Fiore13} study the dispersion induced spectral variations assuming a simple
shape for the GRB intrinsic lightcurve. They assume a gaussian shape,
which is clearly not an adequate description of most GRBs, but has the
advantage of being a simple symmetric function. Furthermore, it
describes sufficiently well at least the core of the main peaks of LAT
GRBs. They further assume that the intrinsic GRB
spectrum, at least in the 100 MeV - 100 GeV range, is well represented
by a single power law model $F(E)=c_{\gamma}E^{-\Gamma}$ photons
s$^{-1}$ cm$^{-2}$ GeV$^{-1}$.
The time resolved spectrum observed at a given time $t^\prime $ will
be related to the intrinsic GRB light curve and to the intrinsic
spectrum as follows:
$$F(t^\prime ,E^\prime )\propto E^{-\Gamma}exp[-{(t-t_0-{E~D\over cE_{QG}})^2\over 2\sigma^2}] ~~~~(3) $$
where $E=E^\prime (1+z)$, and $t=t^\prime /(1+z)$ are the energy and
time in the GRB frame, $D$ is given by eq. 2 and $E_{QG}$ is a quantum
gravity energy scale.
The shape of the observed spectrum is complex
and it will change with the time as a function of $z, E_{QG}$ and
$\sigma$, i.e. burst distance and lenght in addition to the quantum
gravity energy scale.
The rate of variation of the {\it observed} photon index
scale with the inverse of $E_{QG}$ and the inverse of the square of
the GRB duration (for a gaussian shape $T90=4.292(1+z)\sigma$).
As an example, Fiore et al 2013 computed the time resolved ex-
pected spectra assuming a burst at z=3 of gaussian shape,
with $\sigma = 1 $ sec, and $\Gamma = 2.0$. Fig. 2 shows the time re-solved spectra in intervals of 1.3 sec in the observed frame.
\begin{figure*}
\centering
\begin{tabular}{cc}
\includegraphics[width=8cm, angle=0]{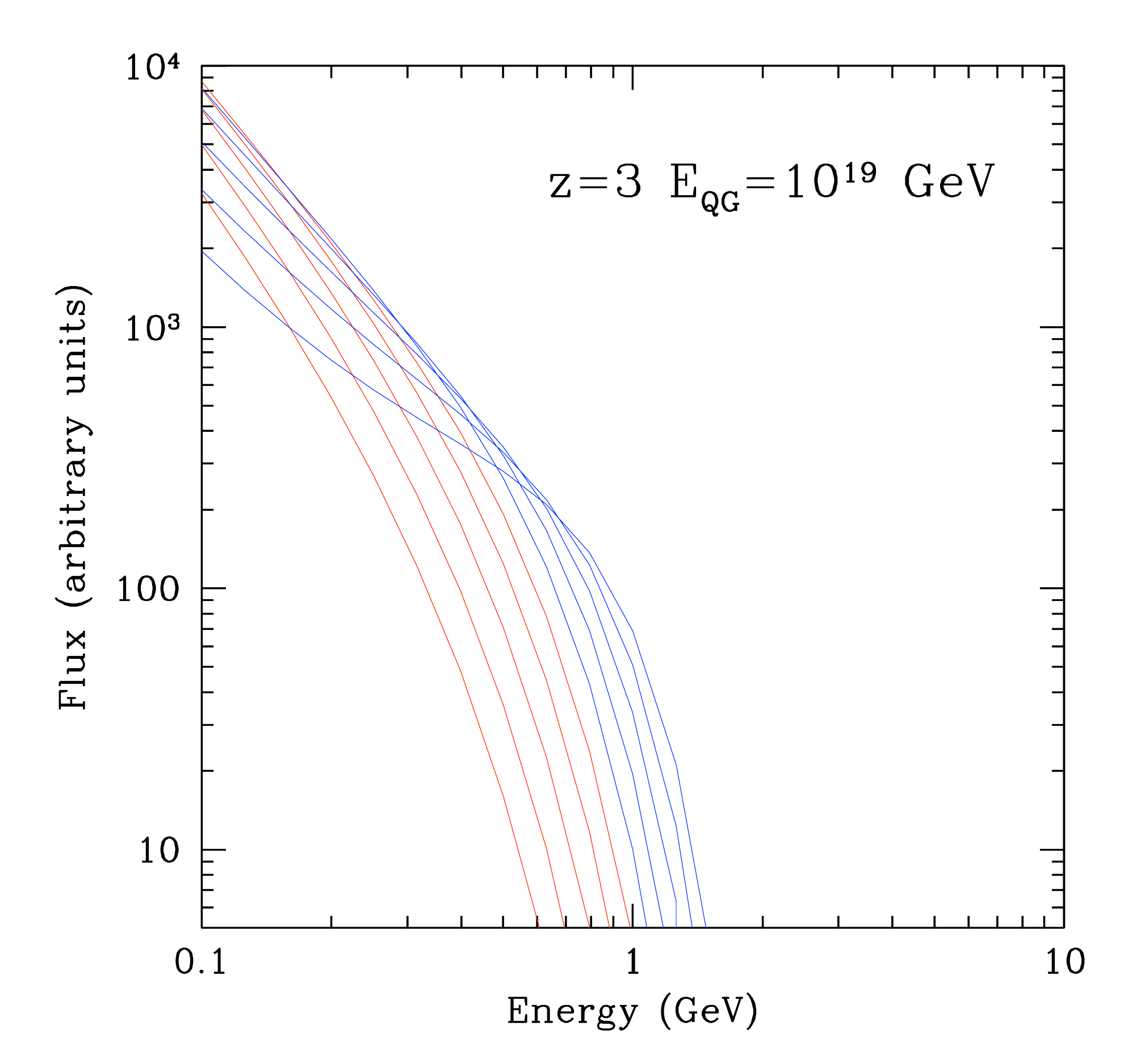}
\includegraphics[width=8cm, angle=0]{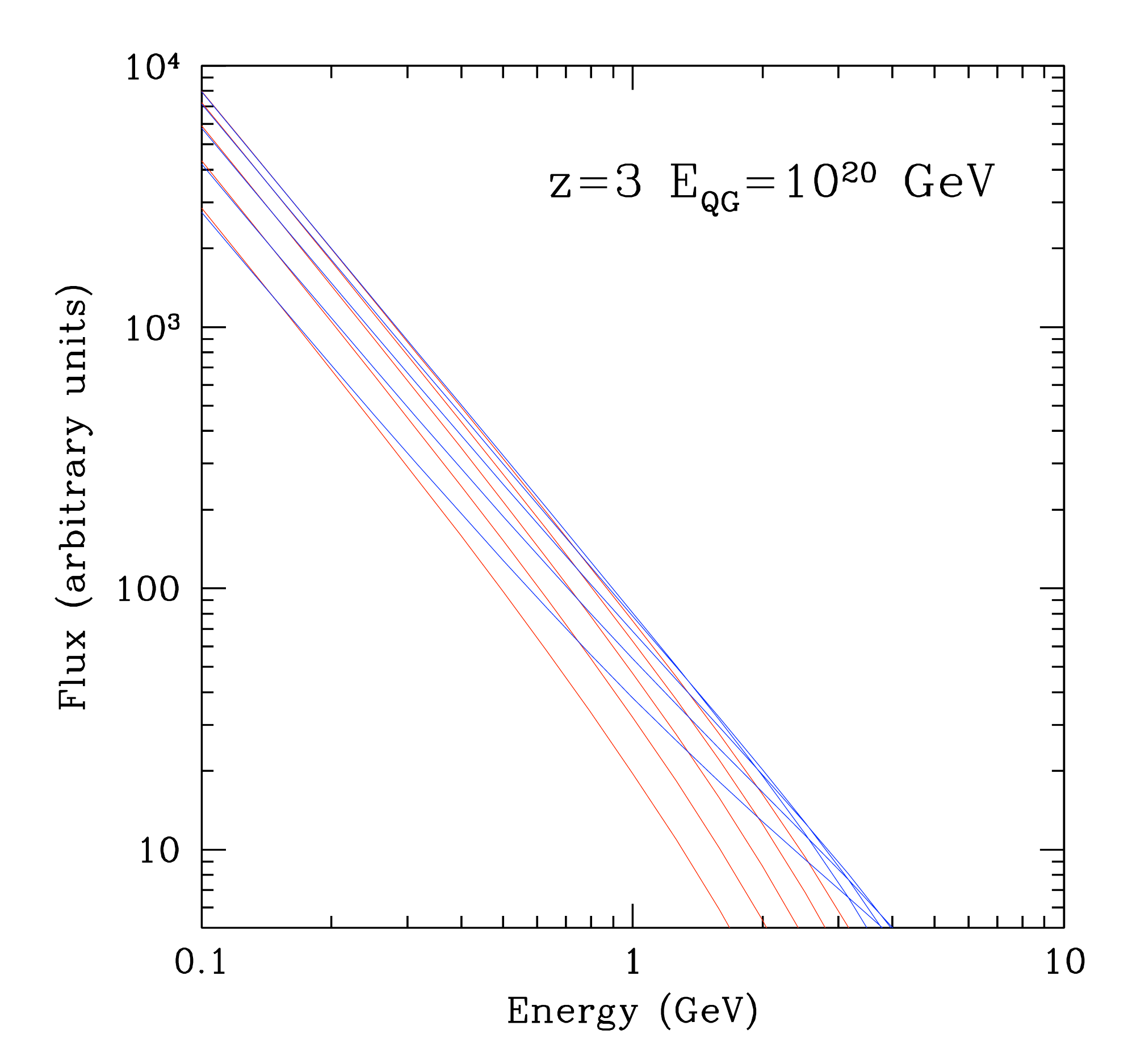}
\end{tabular}
\caption{ Time resolved spectra of GRB with gaussian shape
  ($\sigma=1$ sec, rest frame) and an intrinsic power law spectrum with
  $\Gamma=2$. Light travel effects are computed using eq. 1 and
  assuming that hard photons are lagging softer ones. Red curves
  identify spectra during the rise of the burst and blue curves
  identify spectra during the decrease of the burst. Left panel shows
  ten spectra integrated for 1.3 sec for a burst at z=3 assumuming
  log$E_{QG}=19$. Right panel shows ten spectra
  integrated for 1.3 sec assuming log$E_{QG}=20$. Light travel time
  effects induce a strong spectral hardening with the time and the
  development of a high energy cut-off with energy increasing with
  time. Opposite behaviour (spectral softening and cut-off moving
  from high to low energies) is expected for the case when hard
  photons are travelling faster than softer ones.}
\label{simspe}
\end{figure*}
 Dispersion
effects produce in the observed spectra three main features:(1) a variation of the observed photon index $\Gamma ^\prime$;
(2) a strong high energy cut off with energy changing with the
  time as the GRB develops; (3) the spectrum becomes more complex at GRB observed times
  comparable with the time delay of the high energy photons, with a
  broad peak developing just before the cut-off. This peak is due to
  high energy photons emitted at earlier times, when the burts was
  brighter, but detected at times when the burst faded away.
If the high energy photons are delayed with respect to the low energy
photons (as in the case of Fig.2) the spectral photon index
becomes flatter as the GRB develops, and the cut-off moves from low to
high energies.  Both effects are due to the fact that high energy
photons are systematically shifted at later time. The revers applies
for the case of high energy photons travelling faster than low energy
photons. 
\begin{figure}[ht]
\centering
\includegraphics[width=75mm,height=65mm, angle=0]{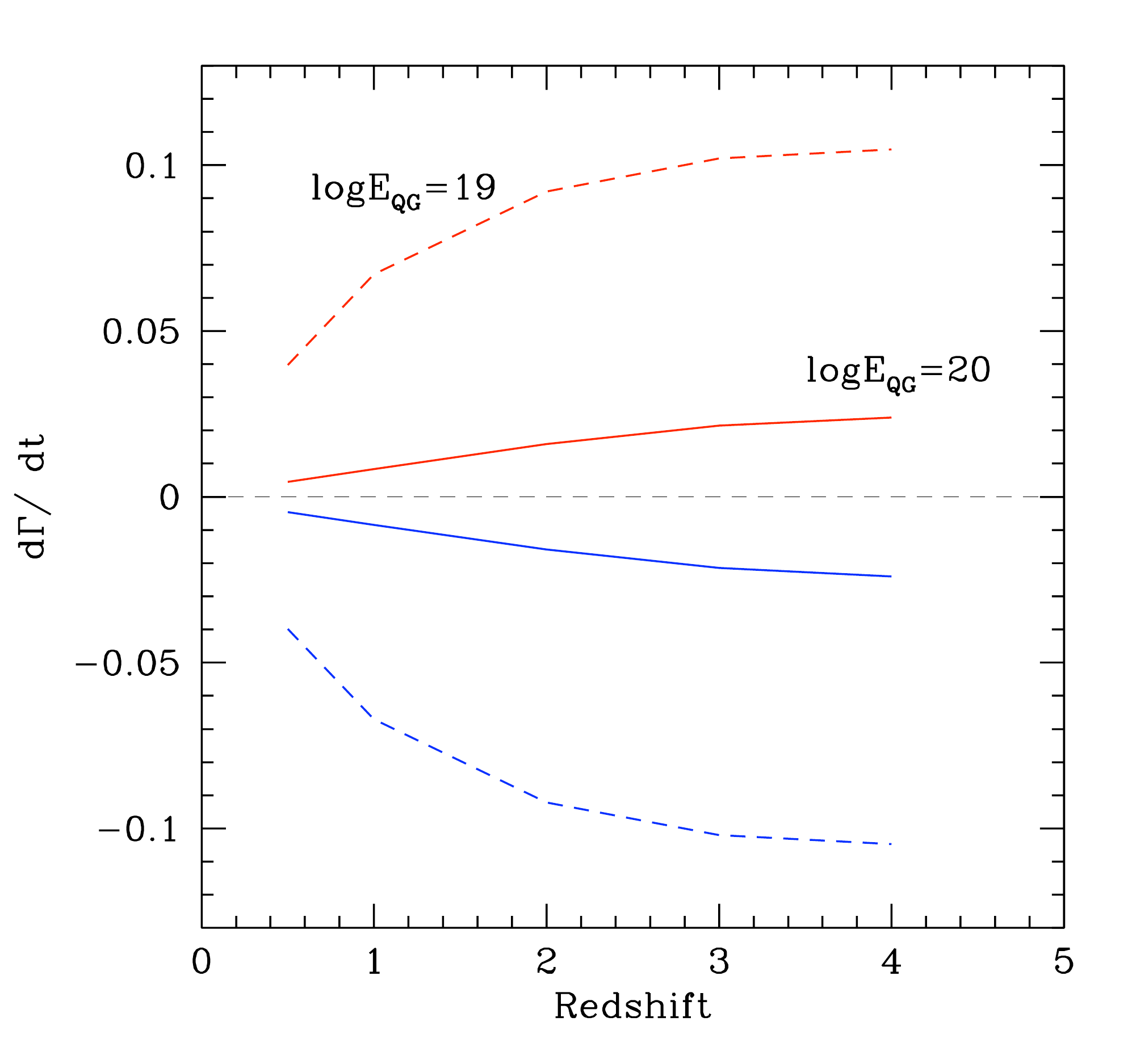}
\caption{ The observed photon index rate of variation ($d\Gamma /dt$) of GRB of
  gaussian shape ($\sigma=2$ sec, observed frame) and intrinsic power law
  spectrum with $\Gamma=2$ as a function of the redshift. Dashed
  curves are the expectation of models with log$E_{QG}$=19 and
  $\alpha=1$, solid curves are the expectation of models with
  log$E_{QG}$=19 and $\alpha=-1$, Negative $d\Gamma / dt$ values refer
  to the case of high energy photons lagging lower energy ones,
  positive values refer to the case of high energy photons preceeding
  lower energy one.}
\label{dgammaz}
\end{figure}
Fiore et al. 2013 approximated  the GRB spectra with the following model:
$$F(E)=c_{\gamma}E^{-\Gamma}exp(-(E/E_{cut})^2)$$
This model is reasonably good approximation for most spectra in
Fig. 2.  As an example, Fig. 3 shows
$d\Gamma ^{\prime}/ dt$ as a function of the redshift for gaussian bursts with
$\sigma=2$ sec (observed frame), $E_{QG}=10^{19}$ GeV and for
$E_{QG}=10^{20}$ Gev.  This analysis shows that $d\Gamma ^{\prime}/ dt$
increases with z quite rapidly and saturates at redshift z=2-3 implying that the QG effect may
be appreciated if LAT will be able to detect several GRBs in the  $0<z<3$ redshift range.

Fiore et al 2013 have applied the method presented above to four bright LAT GRB with a measured redshift
(GRB080916C, GRB090510, GRB090902B and GRB090926A).
These all have well defined peaks lasting up to a few seconds, which 
allow them to study spectral variations on seconds (or fraction of
seconds) timescales.  Spectral variations are detected for GRB090902B
(a spectral hardening with time).  Spectral variations are also
possible in GRB090510 (a spectral softening with time) but not
statistically significant. Previous limits on $E_{QG}$ were obtained using
the arrival time of single high energy photons. In this case limits
are computed by comparing the arrival time of the high energy photon
with that of the the burst onset or that of the GRB peak, because of
course it is not possible to observationally assess the simultaneity
of the high energy photon with some of the low energy photons emitted
at the same time. The computed time delay can be much higher that the
real one, also because it is not easy to uniquely associate one high
energy photon to the GRB prompt event or to its afterglow, thus
decreasing the magnitude of the limits to E$_{QG}$.  Conversely, Fiore et al. 2013
use the full spectrum (typically 100 gamma-ray photons), constraining
its temporal spectral variations on a {\it few second} timescale, thus
avoiding long time delays leading to a potential better sensitivity to the  E$_{QG}$ constraints.
\section{Fermi synergy with other detectors}
The detection of high energy photons from Fermi is not enough to constrain hadronic emission models. The detection of photons (that can be absorbed at the source or on the way from the source to us) or high energy protons (who lose information on the originating source on their way from the source to us) cannot give us enough information on the content of hadrons in the jet and what is their role in the emission mechanism.  Only  the detection of neutrinos that, unlike high energy photons and protons, can travel cosmological distances without being absorbed or deflected can provide information on astrophysical sources that cannot be obtained with high energy photons and charged particles. If hadrons are responsible of the $\sim$GeV emission detected in several GRBs, then this emission is expected to be correlated to the neutrino flux which depends on the energy fraction of protons in the GRB jet. The IceCube upper limits and the Fermi data can be used to constrain the hadronic emission models.
IceCube, completed in December 2010, is the first kilometer-scale neutrino detector. It consists of more than 5000 optical sensors installed at depths from 1,450 m to 2,450 m near the geographic South Pole, over an area of 1 km$^2$. IceCube analyses include a model-independent search for GRB neutrinos\cite{iceCube}, and for other diffuse and point sources. Recent efforts to detect higher energy neutrinos from sources outside our solar system yield important constraints on point-sources and diffuse fluxes of possible sources. IceCube and Fermi overlap in time, therefore the synergy between telescopes is mandatory to constrain the emission mechanism models in GRBs

The GBM has detected several short GRBs. Double neutron star or black
hole/neutron star mergers are believed to give rise to short GRBs. 
Merging binary systems consisting of two collapsed objects are among the most promising sources of
high frequency gravitational wave, signals for ground based interferometers
Short Gamma-Ray Bursts might thus provide a powerful way to infer the merger 
rate of two-collapsed object binaries\cite{GS}. The synergy between GBM and future gravitational
waves detectors like advanced LIGO (ALIGO) and advanced Virgo will be fundamental
to constrain the short GRB progenitor model.
ALIGO will either detect gravitational waves in
coincidence with GBM detections of short GRBs, or neutron star-black  mergers may be ruled out
as the progenitors of these events.

\section{Conclusion}
GeV emission from GRBs is as of yet relatively unexplored observationally. Up to September 2012,
only $\sim$ 35 bursts were detected by LAT in the GeV range and upper limits were put on the GBM GRBs
not detected by LAT\cite{G11,Ben11,Fermi5}.  
 These limits have allowed us to give some constraints on several emitting mechanisms. 
For example,  models where the prompt $\sim 1 $ MeV
emission is produced by inverse-Compton scattering of optical
synchrotron photons  can be ruled out.
The hadronic emission models proposed to explain the $\sim$ GeV emission detected by LAT
can be constrained using the upper limits on the neutrino flux from GRBs recently
reported by the IceCube collaboration\cite{IceCube}. Therefore a synergy between 
Fermi and IceCube is mandatory and only a multiwavelength analysis will allow us to further constrain
the emission mechanism models.

The suppression of the
$\sim100$ MeV flux, compared to that expected from an extrapolation
of the $\sim$1MeV power-law spectrum, suggests that either the
electron energy distribution does not follow a power-law over a
wide energy range, or that the high energy photons are absorbed,
probably by pair production. Requiring an optical depth of $\sim1$
at 100 MeV sets an upper limit to the expansion Lorentz factor
$\Gamma\ls 10^{2.5}[(L/10^{52} {\rm erg/s})/(tv/10 {\rm ms})]1/6$ (e.g. Eq. (7) of\cite{W03}).

Regarding the possibility to constrain the Quantum Gravity effects using the Fermi data
Fiore et al. 2013, presented a method to obtain limits (or detections) to
energy-dependent dispersion effects in Fermi LAT GRBs which uses the
full GRB energy distribution, i.e. the spectrum, and its temporal
variations.  This method provides statistically sounds
confidence intervals (or limits) for the main parameters governing the
energy-dependent dispersion. In order to test this method the ideal would be
to have a bigger sample of LAT GRBs with enough counts in $0.2<z<3$.

In conclusion Fermi has allowed a significant progress in the understanding of the GRB phenomenology,
however some effort should be done in increasing the LAT sensitivity and FOV in order to detect more GRBs at high energy.

\bigskip 
\begin{acknowledgments}
The author wishes to thank the FERMI collaboration for the invitation to give this Review Talk and Giovanni Amelino-Camelia
for useful discussions and comments.
\end{acknowledgments}

\bigskip 

\end{document}